High pressure effects on the electrical resistivity behavior of the Kondo lattice, YbPd2Si2

T. Nakano[1], M. Hedo[1], Y. Uwatoko[1], and E.V. Sampathkumaran[1,2]

[1]*The Institute for Solid State Physics, The University of Tokyo, 5-1-5 Kashiwanoha, Kashiwa, Chiba 277 8581, Japan*

[2]*Tata Institute of Fundamental Research, Homi Bhabha Road, Colaba, Mumbai 400005, India.*

**Abstract**

We report the influence of external high-pressure (P= up to 8 GPa) on the temperature (T) dependence of electrical resistivity ($\rho$) of a Yb-based Kondo lattice, YbPd$_2$Si$_2$, which does not undergo magnetic ordering under ambient pressure condition. There are qualitative changes in the $\rho$(T) behavior due to the application of external pressure. While $\rho$ is found to vary quadratically below 15 K (down to 45 mK) characteristic of Fermi-liquids, a drop is observed below 0.5 K for P= 1 GPa, signaling the onset of magnetic ordering of Yb ions with the application of P. The T at which this fall occurs goes through a peak as a function of P (8 K for P= 2 GPa and about 5 K at high pressures), mimicking Doniach's magnetic phase diagram. We infer that this compound is one of the very few Yb-based stoichiometric materials, in which one can traverse from valence fluctuation to magnetic ordering by the application of external pressure.

The investigation of the competition between magnetic ordering and the Kondo effect continues to be an interesting topic in the field of strongly correlated electron systems. Though both the energy scales characterizing these two phenomena ($T_{RKKY}$ and $T_K$ respectively) depend on the s-f coupling strength (J), the actual functional dependencies are different ($T_{RKKY} \propto (JN(E_F))^2$ and $T_K \propto \exp(-1/JN(E_F))$, where $N(E_F)$ is the density of states at the Fermi energy). These functional forms demand that, for low J values, magnetic ordering dominates, whereas for larger J values, the Kondo effect takes over, with the features attributable to both the phenomena seen prominently for intermediate values of J in the experimental data. As a result of this competition between these two phenomena, there is a maximum [1,2] in the plot of magnetic ordering temperature versus J (well-known as "Doniach's magnetic phase diagram"). Ce and U alloys near magnetic instability (quantum critical point, QCP, in the intermediate range of J) in this plot have attracted a lot of attention in recent years, as Fermi-liquid behavior in the physical properties breaks down in such materials [3]. It is customary to increase J values by applying external pressure (P) or by positive chemical pressure in the case of Ce alloys to traverse from magnetic ordering end to intermediate-valence (IV) end (strong Kondo effect). However, in the case of Yb systems, J decreases with increasing P as 4f-localisation is favored in sharp contrast to the situation in Ce systems. As a result, $T_K$ is bound to decrease for Yb systems with increasing pressure. It is found to be true experimentally [4] and magnetic order could be induced by external P in some Yb-based IV systems, $Yb_2Ni_2Al$ and $YbCu_2Si_2$ [5,6]. There are also a few reports on the positive and also negative chemical-pressure-induced non-Fermi liquid (NFL) anomalies among Yb systems [7-10]. It should however be noted that such studies (that is, evolution of magnetic order from IV state induced by external P) are generally rare among Yb systems. This is because of the fact that one has to choose the Yb Kondo compounds which do not order magnetically (referred to as "non-magnetic Kondo lattices") as 'starting systems', and such compounds are not abundant in number. It is therefore of interest to search for such Yb compounds and subject these to high-pressure experiments. At this juncture, it is worth noting that the compound $YbRh_2Si_2$, being a stoichiometric NFL system, is attracting a lot of attention in the current literature [10,11].

In this article, we focus on the compound, $YbPd_2Si_2$ [12,13], crystallizing in the $ThCr_2Si_2$-type tetragonal structure. Temperature dependencies of magnetic susceptibility, x-ray absorption and x-ray photoemission spectra [13] clearly established that this is a fluctuating valent non-magnetic compound with a fractional valence of Yb very close to 3 with a very weak temperature dependence. The electrical resistivity ($\rho$) studies in the range 1.4 – 25 K indicated Fermi-liquid behavior [9,14]. Heat-capacity studies established that this compound is a heavy-fermion with moderate values (about 200 mJ/mol $K^2$) of electronic co-efficient [15,16]. This compound subsequently generated some interest involving many other experimental methods, including NMR [16], inelastic neutron scattering [17], Yb-quadrupolar moment studies [18], Hall effect and magnetoresistance [19]. In addition, this compound attracted a theoretical interest as well; Schlottmann [20] examined the properties of this compound within the framework of the single-ion Anderson model revealing thereby that the Kondo temperature is of the order of 100 K. It is however surprising that this nearly-trivalent non-magnetic Yb compound has not been studied by high P experiments till to date to simulate an interplay between magnetism and the Kondo effect, particularly noting that the isostructural Yb-based compounds, YbCu$_2$Si$_2$ and YbRh$_2$Si$_2$, yielded interesting results, as briefed above [21]. Therefore, we carried out ρ measurements under P, the results of which are reported in the article.

The sample in the polycrystalline form was prepared by arc melting stoichiometric amounts of constituent elements in an atmosphere of argon. In the case of Yb, the small loss during arc-melting due to low vapor pressure was compensated by adding a requisite amount of Yb after first melting. The ingot was then annealed at 800 C for 1 week and found to be single phase by x-ray diffraction. The ρ measurements were performed at 0, 2, 4, 5, 6, and 8 GPa in a cubic anvil pressure cell in a hydrostatic pressure medium (mixture of fluorinert of FC70 and FC77) in the temperature (T) interval 2.5 – 300 K by a conventional four-probe method employing silver paint to make electrical contacts; in addition, the measurements were extended to 45 mK in a dilution refrigerator at P= 0 and 1 GPa in a piston-cylinder cell employing Daphne oil as a pressure transmitting medium.

We first make some remarks on the ρ(T) behavior for P= 0 (figure 1). ρ decreases nearly linearly down to 175 K below which the decrease is relatively steeper. It was earlier reported [9] that this steepness around 100 K gets reduced by a partial substitution of La for Yb and a minimum appears in the plot of ρ(T) for such alloys, on the basis of which it was concluded that the drop below 100 K represents the onset of coherence among the Kondo-centres of Yb. In the present investigation, we have extended these measurements (under ambient P) to 0.5 K employing a Physical Property Measurement System (PPMS) (Quantum Design) and we find that ρ varies quadratically with T, thereby confirming Fermi-liquid behavior below 1.4 K as well (shown in Fig. 1). We would like to stress that we do not observe any sharp drop down to 45 mK (see below), ruling out any phase transition under ambient pressure conditions and therefore the feature reported at 1.4 K in Ref. 19 must not be intrinsic. A very large magnetoresistance below 30 K, for instance about 25% at 4.2 K for a field of 60 kOe, was also reported in Ref. 19; we have also measured ρ as a function of magnetic field with PPMS, but we found that the magnetoresistance is negligibly small (<<1%) in the entire range of T investigation at such high fields.

We now discuss the T-dependence of ρ at various P, measured in a cubic anvil cell (figure 1). With increasing P, the absolute values of ρ (in the T-range, for instance, above 50 K) decrease monotonically, which indicates that the incoherent Kondo scattering contribution is diminishing. As the T is lowered, the values are higher, say, below 25, 12, and 6 K for 2, 4 and 5 GPa respectively, compared to those at ambient pressure. Thus the ρ-T plots for these pressures intersect with that of ambient pressure at these temperatures. This implies decreasing importance of Kondo coherence effects at low temperatures with increasing P. To illustrate in a different manner, the sharp fall around 100K is absent for higher pressures and ρ decreases relatively smoothly with T. A careful look at the curve for 2 GPa reveals that there is a weak upward bulge around 25 K, as though the Kondo coherence effects set in. With a further increase of P, say, to 4 GPa, this bulge vanishes as though the coherence effects may have fallen below, say, 10 K. All these findings viewed together imply that the Kondo coherence temperature decreases with increasing P. This is in conformity with earlier findings on other Yb systems [4].

We now look at the low temperature behavior (Fig. 2 and 3) more carefully. The 2GPa-plot (Fig. 2), obtained from the cubic anvil cell, actually can be viewed as a nearly linear region in the range 10 to 30 K, followed by a gradual increase in the slope of the $\rho$-T plot as the T is lowered to 5 K; the plot appears to be nearly linear below 5 K again. The intersection point of the lines of these two linear regions is about 8 K. Two alternate explanations can be given to this temperature: (i) Onset of the Kondo coherence effects in the vicinity of 8 K, or (ii) onset of a magnetic transition. In order to understand this better, we show the results of piston-cylinder apparatus extended down to 45 mK at 0 and 1 GPa in figure 3. It is obvious from figure 3 that a drop below 0.5 K for P= 1 GPa appears. A Kondo-coherence can not account for this drop at this temperature, as such a feature should appear between respective temperatures (between 10 and 100 K) for 2 and 0 GPa. Therefore, this drop has a different origin. Since the values of $\rho$ do not go to zero at low temperatures, we at present attribute this drop to the onset of a magnetic transition, rather than to the phenomenon of superconductivity. These findings may imply that that QCP, if it exists, is in the vicinity of 0 to 1 GPa. Now turning to figure 2, as the pressure is increased, say to 4 or 5 GPa, the fall occurs around 5 K. With a further increase of P, this feature remains at 5 K, however preceded by a negative temperature coefficient of $\rho$ in the range 6 – 12 K, that is, a clear minimum around 12 -15 K. The peak close to 5 K and the minimum around 15 K are so well-defined in the plot for 8 GPa that we are tempted to attribute these features to magnetic ordering and the single-ion effect (from the crystal-field-split ground state) respectively. Thus, at this P, the features could be interpreted in terms of an interplay between magnetic ordering and the single-ion Kondo effect. The fact that the upturn due to the single-ion Kondo effect is not seen for P= 4 and 5 GPa could be due to an interference from the onset of a downturn arising from Kondo coherence effects below 10 K at these pressures. In any case, the low temperature transport behavior of this compound under pressure presents an interesting scenario. Incidentally, if the fall at high pressures is truly due to magnetic ordering, the value of the magnetic ordering temperature is rather high for Yb compounds (generally below 5 K); for comparison, the magnetic ordering induced by P for the isostructural IV compound, $YbCu_2Si_2$, occurs around 5 K [6].

To conclude, the application of high pressure on $YbPd_2Si_2$ results in remarkable changes in the electrical resistivity behavior. A drop in $\rho(T)$ appears at low temperatures for all externally applied pressures. Since the $\rho$ values do not attain zero, we believe that this drop originates from magnetic ordering, rather than from superconductivity – hitherto unknown among Yb systems (unless one invokes the idea of filamentary superconductivity). The temperature below which the fall in $\rho$ occurs goes through a peak with P (0.5, 8, 5, 5, 5, and 5 K for P= 1, 2, 4, 5, 6, and 8 GPa), as though the magnetic ordering temperature goes through a maximum around 2GPa in the Doniach's diagram. While we call for a conclusive evidence for magnetic ordering under pressure from other experiments, e.g., high pressure $^{170}$Yb Mössbauer effect and magnetization measurements, we hope that this high pressure work will stimulate more investigations on this compound to understand competition between magnetism and the Kondo effect among Yb systems and to probe QCP effects.

21. For thermal expansion anomalies under pressure in YbCu$_2$Si$_2$, see, Y. Uwatoko, G. Oomi, J.D. Thompson, P.C. Canfield, and Z. Fisk, Physica B **186-188**, 593 (1993).

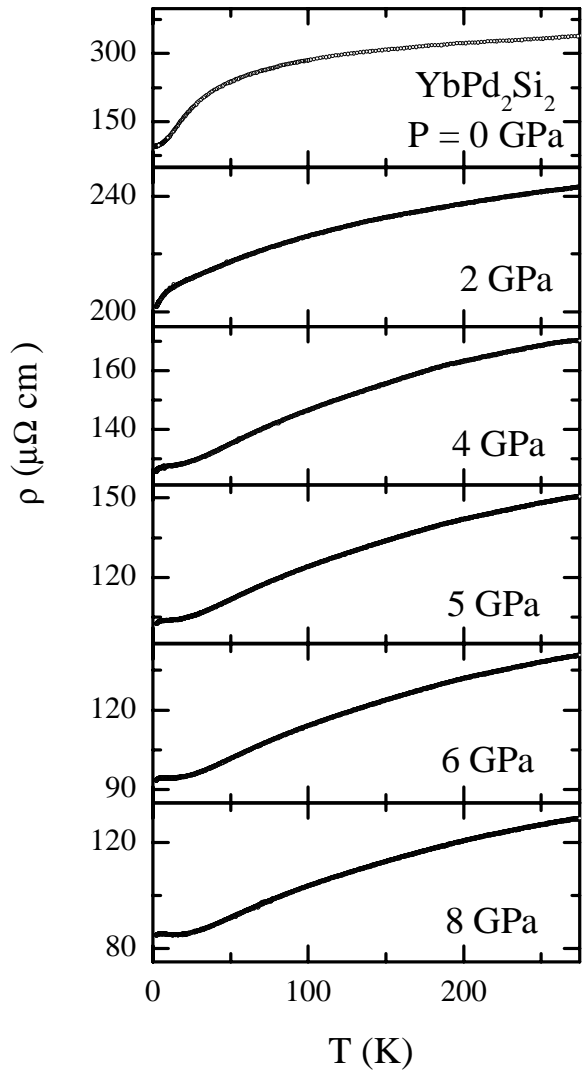

Figure 1:
Electrical resistivity ($\rho$) behavior over a wide temperature range for YbPd$_2$Si$_2$ at various external pressures, obtained from a cubic-anvil pressure cell.

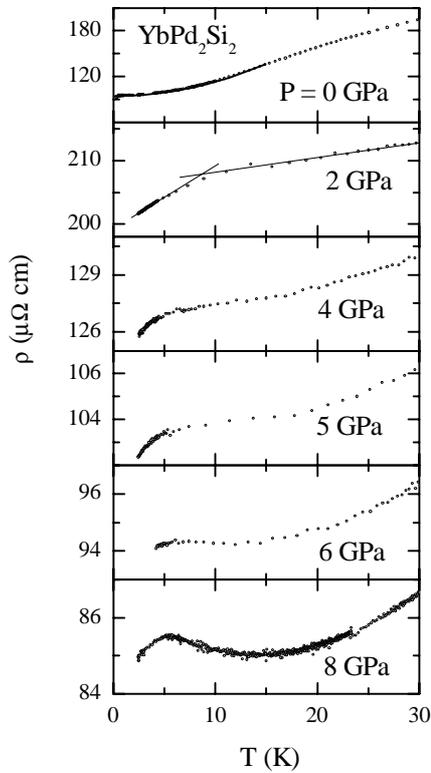

Figure 2:

Electrical resistivity (ρ) behavior below 30 K for YbPd$_2$Si$_2$ at various external pressures (P), obtained from a cubic-anvil pressure cell. For P= 0, the continuous line is obtained by a fit to the quadratic temperature dependence of the data in the range 0.5 to 15 K. For 2 GPa, the two regions, described in the text, are shown by continuous lines.

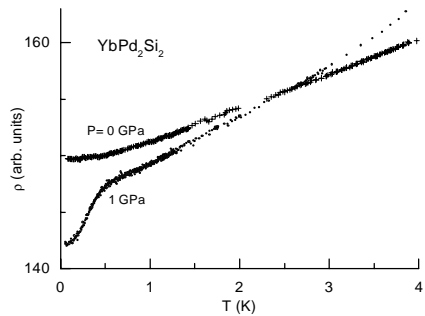

Figure 3:

The low temperature (down to 45 mK) electrical resistivity behavior of YbPd$_2$Si$_2$ obtained from a piston-cylinder pressure cell in a dilution refrigerator for P= 0 and 1 GPa.